# Evidence of Indium impurity band in superconducting (Sn,In)Te thin films

Jiashu Wang[1], Trisha Musall[1], Bo-An Chen[2,3], Marie Gerges[1], Sylwia Ptasinska[1,2], Xinyu Liu[1], Badih A. Assaf[1]

[1] Department of Physics and Astronomy, University of Notre Dame, Notre Dame IN, USA

[2] Radiation Laboratory, University of Notre Dame, Notre Dame, IN 46556, USA

[3] Department of Chemistry and Biochemistry, University of Notre Dame, Notre Dame IN, USA

**Abstract**: $Sn_{1-x}In_xTe$ has been synthesized and studied recently as a candidate topological superconductor. Its superconducting critical temperature increases with Indium concentration. However, the role of Indium in altering the normal state band structure and generating superconductivity is not well-understood. Here, we explore this question in $Sn_{1-x}In_xTe$ (0<x<0.3) thin films, characterized by magneto-transport, infrared transmission and photoemission spectroscopy measurement. We show that Indium is forming an impurity band below the valence band edge which pins the Fermi energy and effectively generates electron doping. An enhanced density-of-states due to this impurity band leads to the enhancement of $T_c$ measured in multiple previous studies. The existence of the In impurity band and the role of In as a resonant impurity should be more carefully considered when discussing the topological nature of $Sn_{1-x}In_xTe$.

**Introduction**

Topological superconductors (TSCs) are superconductors that have superconducting gap in the bulk with gapless boundary states. They not only enrich scientists' understanding of topological phases, but also have promising applications. [1–3] Chiral p-wave topological superconductors for example, are potential platforms for topological quantum computing [4]. These materials are rare in nature. There are ongoing efforts to artificially engineer topological superconductors using semiconductor-superconductor heterostructures. [5] Other ongoing efforts also aim to find intrinsic topological superconductors. In the recent decade, the following candidate materials have been proposed as potential topological superconductors: $Cu_xBi_2Se_3$ [6,7], $UTe_2$ [8], $UPt_3$ [9] and $Sn_{1-x}In_xTe$ [10,11].

SnTe is a topological crystalline insulator (TCI). [12,13] It hosts four Dirac cones protected by mirror symmetry on its (111) and (100) surfaces. Through indium doping, it becomes superconducting without losing its topological properties. [10,14] It is a TSC candidate, since its superconducting state can be described by the Hamiltonian proposed by Fu and Berg [6,10] to predict the TSC state in $Cu_xBi_2Se_3$. [11] The $Cu_xBi_2Se_3$ material family has been well studied experimentally. There, Cu intercalates between the Se layers and acts as an electron donor. Its role in generating superconductivity is understood. [7,15] The pairing symmetry in $Sn_{1-x}In_xTe$ remains controversial, [16–18] but recent work has shown it is dominantly s-wave [19] at high

In-content, despite previous work arguing that surface states persist in the superconducting state [16]. The role of In in $Sn_{1-x}In_xTe$ still lacks sufficient understanding. In is usually considered an acceptor in IV-VI materials [14] but prior studies on single crystals have argued that In actually forms an impurity band near the Fermi level and acts effectively as a donor. [20–22] One manifestation of the impurity band in these studies, was the reduction of the Hall constant, associated with coexisting electrons and holes. A spectroscopic measurement is needed to probe the band structure and to provide evidence of the existence of the impurity band.

Here, we analyzed the role of In in $Sn_{1-x}In_xTe$ thin films by a combination of magnetotransport, infrared measurements, and X-ray photoelectron spectroscopy. Magneto-transport measurements show that the Hall slope of $Sn_{1-x}In_xTe$ films becomes 100 times smaller with increasing In concentration and increasing $T_c$, consistent with prior work on single crystals. The calculated carrier density is however unreasonably high, indicating a likely existence of both electrons and holes. We consider the impurity band model to explain this observation, and we corroborate the model using optical measurements of the absorption edge that evidence a pinning of the Fermi energy in the valence band. Lastly, X-ray and ultraviolet photoemission spectroscopy (XPS and UPS) measurements confirm that In is indeed a resonant impurity and introduces a *d*-like band below the valence band edge of SnTe. Our studies show that pairing involving indium impurity states should be taken into consideration carefully when discussing the topological properties of $Sn_{1-x}In_xTe$, especially at high In content needed to enhance $T_c$ above 2K.

**Experiments**

1. Sample growth

Superconducting $Sn_{1-x}In_xTe$ films on $BaF_2$ (111) substrates with controlled Indium concentration(x=0-0.3) are grown by molecular beam epitaxy. A 25nm SnTe layer is grown first, followed by a 75nm $Sn_{1-x}In_xTe$ layer. We find that the indium diffuses fully into the SnTe layer, thus forming a uniform 100nm $Sn_{1-x}In_xTe$ single layer (Fig. 1(a)). X-ray diffraction (XRD) analysis published in our previous work show the sample has a rocksalt structure, with slight compressive in-plane strain and tensile strain in the out-of-plane direction. The In concentration, sample thickness, and strain are confirmed by XRD, transmission electron microscopy (TEM) and energy dispersive spectroscopy (EDS). The characterization results are shown in ref. [23].

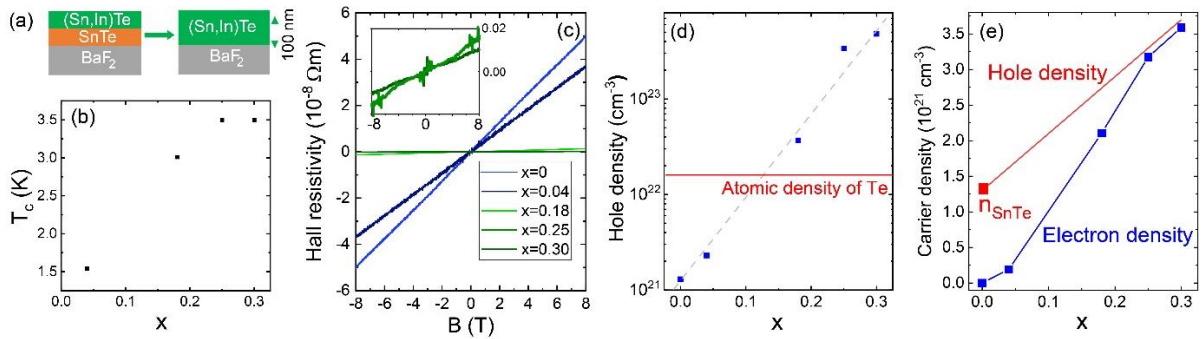

**Figure 1**. (a) Sample structure. 25nm SnTe and 75nm $Sn_{1-x}In_xTe$ are grown in sequence on $BaF_2$ (111) substrate. In diffuses into the SnTe layer, resulting in a uniform $Sn_{1-x}In_xTe$ layer. (b) Critical superconducting temperature $T_c$ of samples with different In content x. (c) Hall resistivity of samples with different In content x. Inset: Zoom-in image of the Hall resistivity for x=0.25 and x=0.3. (d) Extracted carrier density, assuming a single-carrier type (p-type). The gray dashed line is a guide for the eye. The red line is the calculated atomic density of Te, assuming the lattice constant is 6.3Å. (e) Extracted electron density with two-carrier model assuming electrons and holes with similar mobility coexist and hole density increases linearly with In concentration x. The red dot is the experimental data of pure SnTe film.

## 2. Magnetotransport Measurements

Electrical magnetotransport measurements are performed between 1.5K to 15K in a custom system equipped with a standard superconducting, covering both range of superconducting and normal state. The critical superconducting temperature $T_C$ increases with x, and can reach 3.5K with x=0.3 (Fig. 1(b)). The order of magnitude of $T_c$ is consistent with pervious work on thin films of (Sn,In)Te. [24,25] The superconducting properties, as well as the longitudinal magneto-resistance in the normal state, are discussed in detail in ref. [23]. Here, we will focus on the Hall effect and optical properties of $Sn_{1-x}In_xTe$ in the normal state, as they tightly relate to the role of the In impurity.

The Hall resistivity of our $Sn_{1-x}In_xTe$ films at 6K is shown in Fig. 1(c). All samples exhibit a p-type behavior with a constant Hall slope. As x increases, the Hall slope decreases dramatically. Particularly, for samples with x=0.25 and x=0.3, the Hall slope is two-orders smaller than for those with lower x (Fig. 1(c) inset). If we assume that In is providing acceptors and generating holes, the hole density can be calculated from the Hall slope and given as blue dots in Fig. 1(d). With this assumption, we recover unreasonably large hole densities for high x. From the average lattice constant 6.3Å, the Te atomic density is around $1.6 \times 10^{22} cm^{-3}$, indicated by the red line in Fig. 1(d). Hole densities for $x \geq 0.18$ exceed this limit by more than an order of magnitude.

The most plausible explanation of this result is that a portion of In actually act as electron donors ($In^{3+}$). Meanwhile, the remaining In ($In^{1+}$) as well as Sn vacancies act as acceptors providing holes. The system then hosts coexisting electrons and holes with a similar but low mobility. Under such conditions, the Hall resistivity in the Drude model follows:

$$\rho_{xy} \approx \frac{n_h - n_e}{e(n_h + n_e)^2} B \qquad (1)$$

Where $n_{h,e}$ is the carrier density of holes and electrons. The above equation holds when their mobilities are almost equal, $\mu_h \approx \mu_e$. Such approximation is based on two experimental facts: (1) The Hall slope remains perfectly linear at all Indium concentrations, which indicates carriers having similar mobility; (2) The hole mobility of pure SnTe (x<0.01) grown for this work is already quite low (10cm²/Vs to 59 cm²/Vs). Thus, when the electron density is close to the hole density, the slope of the Hall resistivity is reduced resulting in an abnormally large density of "single carriers". We note that in ref. [20], authors made a similar observation but fit their Hall measurements by considering a changing mobility of localized electrons. Since there are more fitting parameters than constraints in this model, both our conclusions are similar in reflecting the role of electrons in transport. $n_e$ can be viewed as a density of "effective free electrons" with mobility equal to that of holes.

Under these assumptions, we can utilize the hole density $n_0$ of pure SnTe (x=0) to approximate the electron density of other samples. We assume an approximation that the hole density increases linearly with In concentration:

$$n_h = n_{SnTe} + \frac{x}{2} \cdot n_0$$

where $n_{SnTe}$ is the measured hole density of pure SnTe film, $n_0$ is the atomic density of Te. The approximation here is assuming half the Indium acts like $In^{1+}$ and donates holes and the other half acts like $In^{3+}$ and donates electrons. The calculated $n_e$ (effective free electrons) is not necessarily equal to the density of $In^{3+}$ since a portion of the impurity carriers may be localized. $n_h$ is taken into Eq. 1 to calculate $n_e$. The result is shown in Fig. 1(e). As the In concentration rises, more electrons are contributing to the transport, and causing the Hall slope to decrease dramatically.

Previous studies have also suggested that as In concentration gets higher, In-5s and Te-5p orbitals hybridize and form an impurity band below the Fermi level $E_f$. [26] This together with the transport analysis leads us to a physical picture as shown in Fig. 2. With small x, most indium atoms are $In^+$ and thus donate holes to the system bringing $E_f$ further down into the valence band as indicated by the decreasing Hall slope. When In concentration becomes large enough, an In5s-Te5p hybridized impurity band forms and pins the $E_f$ to its position. The impurity band donates electrons to the system.

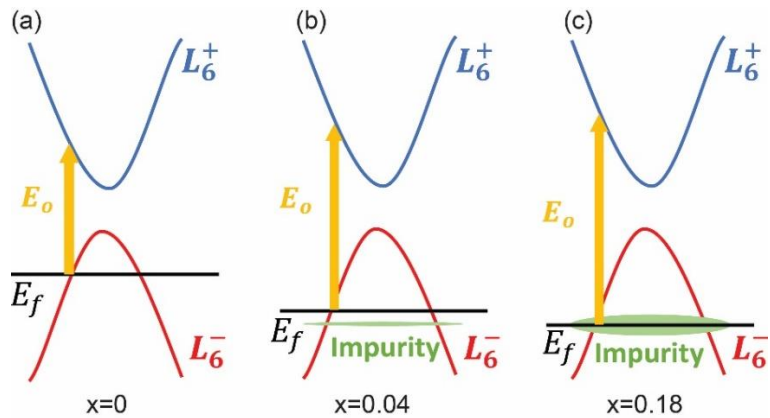

**Figure 2**. Evolution of the band structure with increasing In concentration x. (a) x=0, Fermi level $E_f$ lies below the valence band edge. Yellow arrow indicates the energy of the optical transition (optical gap $E_o$) from the conduction band to the valence band. (b) x=0.04, $E_f$ moves further down, and an indium impurity band forms (green). (c) x=0.18, $E_f$ reaches the impurity band region and gets pinned to it. $L_6^\pm$ denote the conduction and valence band edges of SnTe at the L-points of the Brillouin zone.

## 3. Optical Infrared Spectroscopy

The above scenario is based on magnetotransport measurements and follows what is hypothesized in previous studies using DFT calculations [26]. To validate such a model, we additionally perform Fourier-transform infrared (FTIR) transmission spectroscopy at room temperature in a Brucker Vertex 80v setup equipped with a CaF$_2$ beam splitter for the near-infrared to reach 15000cm$^{-1}$. Our thin films enable measurements in the transmission geometry not possible on thick conducting crystals. By dividing the transmitted intensity through the samples ($T_{sample}$) by that measured through a pure BaF$_2$ substrate ($T_{sub}$), we can get the relative transmission signal through each film. The results are shown in Fig. 3a. A wide transparency region is seen in the mid-infrared well into the near-infrared for most samples, consistent with Sn$_{1-x}$In$_x$Te retaining a semiconducting normal state. A sudden drop in the intensity indicates a corresponding optical inter-band transition with energy equal to the photon energy $h\nu$. It is plotted as the yellow arrow $E_o$ in Fig. 2. The position of this transmission edge blue-shifts with increasing x, indicating an increasing optical gap. The detection of $E_o$ is important as it directly probes the position of the Fermi level $E_f$. That is because Sn$_{1-x}$In$_x$Te is degenerately doped, which Pauli blocks band-edge to band-edge transitions. The measured optical gap will then include the Moss-Burstein shift and is given by $E_o = E_g + 2E_f$, assuming electron-hole symmetry.

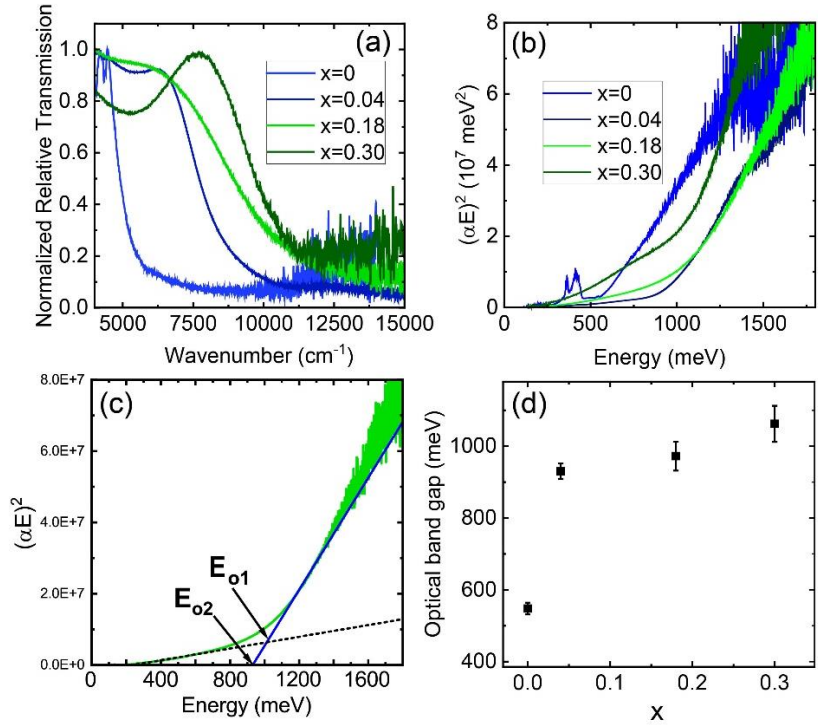

**Figure 3**. FTIR transmission spectroscopy for Sn$_{1-x}$In$_x$Te films with different x. (a) Relative transmission intensity of samples in the mid-infrared and near-infrared range. The drop indicates the absorption caused by inter-band optic transitions. (b) Tauc plot of the data shown in (a). (c) Extraction of the optical gap $E_o$ from the Tauc plot shown for x=0.18. The blue is a linear fit of the absorption edge, and the dashed black line is a linear fir of the background. Two values of $E_o$ are determined from the intercept of the blue line

with the x-axis $E_{o2}$ and with the background $E_{o1}$. The average of the two is reported and their standard deviation is taken as the uncertainty on $E_o$. (d) Extracted optical gap $E_o$ of the four samples versus x.

A more accurate estimation on $E_o$ is to use Tauc plot method, expressed by equation [27]:

$$(\alpha \cdot h\nu)^{1/\gamma} = B(h\nu - E_o)$$

Here $\alpha = -\ln\left(\frac{T_{sample}}{T_{sub}}\right)$ is the absorption coefficient, $\gamma = 1/2$ for direct band transition. By this equation, it becomes more convenient to rearrange the data to a Tauc plot in Fig. 3(b). The intercept between the rising linear absorption edge in the Tauc plot and the x-axis is the value of the optic gap, as illustrated by $E_{o2}$ in Fig. 3(c). Since our baseline is not perfectly flat, we can also extract the optical gap as the point of intercept with an energy dependent baseline, as indicated by $E_{o1}$ in Fig. 3(c). Thus, to include the effect of uncertainty from the background, an average of $E_o$ along an error can be extracted for all samples from the two values $E_{o1,o2}$.

The results of $E_o$ as a function of composition x are shown in Fig. 3(d). The optical gap of SnTe is found to be 550meV. If we take the direct fundamental band gap of SnTe to be $E_g$=0.18eV [28,29], this yields a Fermi level of 175meV below the valence band edge without In. The qualitative behavior of the optical gap with increasing In content is important to highlight. It increases to 930meV for x=0.04, but rises more slowly between x=0.04 to x=0.30, despite the Hall voltage dropping by two orders of magnitude. This is consistent with a pinning of $E_f$ attributed to the emergence of an impurity band.

The trend of optical gap with x agrees with the proposed model in Fig.2. The value of the optical gap $E_o$ directly reflects the position of the Fermi level. Larger $E_o$ implies $E_f$ lies deeper in the valence band. At small x, In mostly gives holes to the system, leading to a fast drop of the Fermi level compared to SnTe. At x>0.04, an impurity band forms and $E_f$ gets pinned at its position. $E_o$ then remains relatively unchanged. If we assume that (Sn,In)Te also has a direct band gap $E_g$ of 0.18eV at room temperature, this indicates that the impurity band lies around 400meV below the valence band edge.

It is worth noting here that in SnTe, a heavy hole band lying 0.3-0.5eV below the valence band edge, has been identified in the past. It is referred to as the Σ band. It needs to be ruled out as possible source of Fermi level pinning that can explain the flat trend of $E_o$ upon introduction of Indium. If $E_f$ crosses the Σ band, the density-of-states at $E_f$ becomes larger and requires more holes to lower $E_f$. However, the slope of $E_o$ vs x in Fig. 3(d) is extremely small and does not exceed 5meV/1%. Additionally, the behavior of the Hall voltage, dropping by almost an order of magnitude, cannot be accounted for by the Σ band. Additionally, taking the heavy hole effective mass to be $m_\Sigma \approx 2m_0$ [30,31] we can compute the hole density supported by this band as a function of energy assuming a simple spherical Fermi surface. This assumption will over-estimate the carrier density. We find that $E_f$ needs to be more than 1.5eV below the Σ-band to get a carrier density that exceed $10^{23}$cm$^{-3}$ which is not consistent with the pinning of $E_f$. Hence, the density-of-states of the heavy hole Σ-band is not sufficiently high to account for our measured Hall density. In ref. [23] we showed that the lattice constant of $Sn_{1-x}In_xTe$ decreases linearly with x. From x=0 to x=0.30 the lattice constant changes quasi-linearly from 6.33Å to 6.27Å. Thus, it is necessary to consider the role of strain effect on the band gap. If we view it as an effective compressive strain to the SnTe lattice, the modification to the band gap can be calculated using:

$$\Delta E_g = D_d(2\varepsilon_\parallel + \varepsilon_\perp) + D_u(8\varepsilon_\parallel + \varepsilon_\perp)/9$$

where $\varepsilon = \frac{a}{a_0} - 1$ and $D_{d,u}$ are deformation potentials which can be found in ref [32]. Here, we are neglecting the strain coming from the BaF$_2$ substrate and assume the x=0 lattice constant is the relaxed $a_0$. Then, we get that the band gap modification is negligible for x=0.04, 29meV for x=0.18 and 35meV for x=0.30. Thus, strain can increase the band gap by no more than 35meV, which can partially explain the increase of the optical gap in x>0.04 region but cannot explain the strong enhancement of $E_o$ from x=0 to x=0.04.

The optical measurements shown in Fig. 3 are all carried out at room temperature, but we also performed low-temperature FTIR measurements down to 4.2K (see appendix A). As $E_f$ drops with temperature, $E_o$ also drops. But sample-to-sample variations due to the changing chemical composition remain much larger than the slight temperature dependence of the gap. We conclude that the discussion of optical measurements at room temperature also applies at low temperature to the normal state of the system right before it goes superconducting.

4. **X-ray and Ultraviolet Photoemission Spectroscopies**

Next, photoelectron spectroscopic measurements are performed using a system which combines two photoemission sources: X-rays and ultraviolet light, purchased from SPECS Surface Nano Analysis GmbH (Berlin, Germany). XPS spectra were recorded using a monochromatized Al Kα (1486.7 eV) X-ray beam generated by a micro-focus X-ray source. While, UPS spectra were recorded using He I (21.2 eV) generated by a plasma excitation source. The details of this experimental system are described previously [33]. The details of the sample preparation are discussed in Appendix B.

UPS data, shown in Fig. 4(a), further reveal the changing shape of the valence band upon introduction of In. Slightly below the Fermi energy, marked by a black arrow, a peak is observed for both x=0.04 and x=0.3. The peak does not shift in energy, but the intensity becomes stronger with In concentration. Such behavior indicates an enhanced density-of-states in the valence band of $Sn_{1-x}In_xTe$, consistent with the impurity band picture. The results of UPS imply that the increasing In leads to the formation of an impurity band below the valence band edge likely due to the In5s-Te5p hybridization.

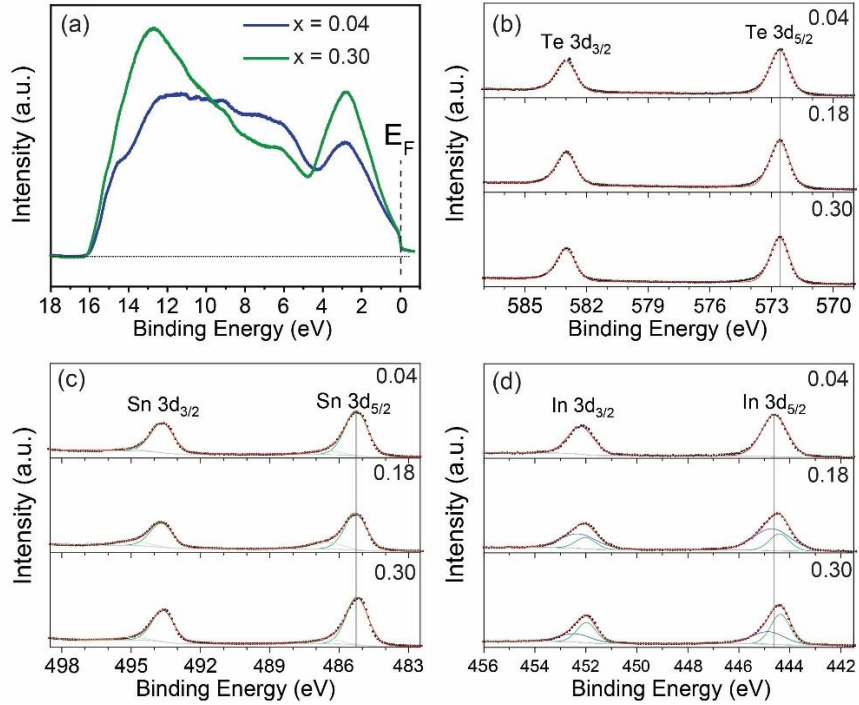

**Figure 4**. (a) UPS spectra of $Sn_{1-x}In_xTe$ films with x=0.04 and 0.30. XPS spectra of the Te-3d (b), Sn-3d (c), and In-3d (d) peaks for x=0.04, 0.18 and 0.30. The Sn spectra are used for calibration of the binding energy. The black dots represent data points and the solid red lines represent the overall fit to the data. The green and blue lines represent individual peaks included in the fit.

Lastly, to further elucidate the mixed valence of Indium needed to explain the Hall effect, we performed X-ray photoelectron spectroscopy (XPS) on the x=0.04, 0.18 and 0.30 samples (Fig. 4(b-d)). All XPS spectra are calibrated using the Sn-$3d_{5/2}$ peak at a binding energy of 485.2eV. [34] The Sn-3d XPS peaks are shown in Fig. 4(c). The Te-3d peaks (Fig. 4(b)) do not exhibit any dramatic changes with increasing In content. The In-3d peaks are shown in Figure 4(d) and analyzed in detail in Appendix 2. We notice that the peak position is almost independent of x (within 0.1eV) from x=0.04 to x=0.30. However, the In-3d peaks become more asymmetric with increasing x, indicating the likely presence of a second peak on the high energy tail of the main one (Fig. 4(d)). In is known to skip the 2+ valence [35] and has been hypothesized to have mixed (+1 and +3) valence in $Sn_{1-x}In_xTe$, similar to Tl in PbTe [36,37]. We cannot however distinguish the two components and attribute them to specific oxidation states, as their peak values are closer to each other than what is expected for $In^{+3}$ and $In^+$. [38] However, the proximity of the $In^+$ and $In^{3+}$ XPS peaks makes it challenging to fit spectra and to confirm their mixed valence character (see Appendix B). It is also possible that the high energy peak is due to a strong In5s-Te5p hybridization reported in density-functional-theory calculations [21]

**Conclusion**

In summary, magnetotransport, optical absorption and photoemission spectroscopy measurements corroborate the presence of an In impurity band in $Sn_{1-x}In_xTe$ thin films.

Experimentally, a dramatic decrease of the Hall slope with increasing In concentration x is observed, while optical measurements reveal a weakly varying optical gap for x≥0.04. These observations indicate that the Fermi level is pinned to an impurity band. The hypothesis is further supported by UPS measurements which show the changing shape of the valence band. While XPS measurements could not conclusively differentiate specific oxidation states of In, they did reveal an asymmetric In peak at high In content. This is consistent with the mixed valence picture, but could also originate from a hybridization of the In and Te levels. Since $Sn_{1-x}In_xTe$ is commonly viewed as a potential topological superconductor, the existence of the In impurity band should be more carefully considered when discussing its topological nature. Prior work hypothesizing topological superconductivity has essentially considered pairing near the valence band edge for this material, involving Sn and Te orbitals of opposite parity [11]. However, the experimental results shown here, as well as prior measurements on single crystals consistent with our interpretation, indicate that such a treatment is not sufficient to understand the nature of superconductivity in this material. [21,23,39,40] At the very least, proper models of the superconductivity of $Sn_{1-x}In_xTe$ must take into account the changing band structure of this materials' valence band upon introduction of In.

**Acknowledgements.** Work supported by National Science Foundation Award DMR-1905277. TM also acknowledges support from the Notre Dame REU program through National Science Foundation Award PHY-2050527. BC and SP acknowledge support from the U.S. Department of Energy Office of Science, Office of Basic Energy Sciences under Award Number DE-FC02-04ER15533.

**Appendix A: Low temperature optical transmission spectroscopy.**

The optical transmission spectroscopy measurements discussed in manuscript are repeated ta low temperature, down to 4.2K. The variation of the optical gap versus temperature is shown in Fig. 5. It is evident that the sample-to-sample variations seen at room temperature remain valid at low temperature.

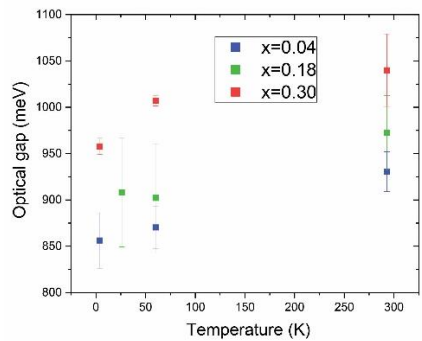

**Figure 5.** Temperature dependence of the optical band gap in $Sn_{1-x}In_xTe$ films.

**Appendix B: Analysis of the XPS spectra.**

XPS measurements are carried out on $Sn_{1-x}In_xTe$ with x=0.04, 0.18 and 0.3, after removal of an oxide layer by $Ar^+$ ion sputtering. The presence of the oxide layer is monitored by measuring the O-1s, Sn-3d and Te-3d signal (4+ oxidation of Te and Sn appear due to $SnO_2$ and $TeO_2$). Sputtering is carried out until we can no longer detect those specific signals. The results shown in Fig. 4 are obtained after the removal of the oxide.

A GL(30) line shape function (70% Gaussian and 30% Lorentzian) is employed to fit individual peaks and a Shirley-type background is defined. The results for Sn and Te respectively confirm the 2+ and 2- oxidation expected for the two elements. The fitting results obtained for the In $3d_{5/2}$ peak are shown in table 1.

|  | x=0.04 | x=0.18 |  | x=0.30 |  | InI | $InI_2$ | $InI_3$ |
|---|---|---|---|---|---|---|---|---|
| Binding energy (eV) | 444.6 | 444.7 | 444.4 | 444.8 | 444.4 | 444.0 | 445.1 | 445.1 |
| FWHM (eV) | 1.18 | 1.7 | 0.91 | 1.53 | 0.88 |  |  |  |
| Weight | 1 | 0.70 | 0.30 | 0.42 | 0.58 |  |  |  |
| Reference | This work $Sn_{1-x}In_xTe$ |  |  |  |  | [41] |  |  |

**Table 1.** XPS spectra fitting results for the curves shown in Fig. 4(d).

Sample to sample variations in the binding energies do not exceed 0.3eV. The shoulder that accounts for the peak asymmetry for x=0.18 and x=0.3 in Fig. 4(d), yields a peak that is at most 0.4eV off the main peak. These variations are smaller than variations seen in the In $3d_{5/2}$ peak as the oxidation state changes from 1+ (InI) to 3+ ($InI_3$). The table shows this comparison for iodides as a reference that this holds for other binary In compounds. [38,41]

[1] X.-L. Qi and S.-C. Zhang, *Topological Insulators and Superconductors*, Rev Mod Phys **83**, 1057 (2011).

[2] X. L. Qi, T. L. Hughes, S. Raghu, and S. C. Zhang, *Time-Reversal-Invariant Topological Superconductors and Superfluids in Two and Three Dimensions*, Phys Rev Lett **102**, (2009).

[3] M. Sato and Y. Ando, *Topological Superconductors: A Review*, Reports on Progress in Physics **80**, 076501 (2017).

[4] S. Das Sarma, M. Freedman, and C. Nayak, *Majorana Zero Modes and Topological Quantum Computation*, Npj Quantum Inf **1**, 15001 (2015).

[5] L. Fu and C. L. Kane, *Superconducting Proximity Effect and Majorana Fermions at the Surface of a Topological Insulator*, Phys Rev Lett **100**, 096407 (2008).

[6] L. Fu and E. Berg, *Odd-Parity Topological Superconductors: Theory and Application to CuxBi2Se3*, Phys Rev Lett **105**, 097001 (2010).

[7] Y. S. Hor, A. J. Williams, J. G. Checkelsky, P. Roushan, J. Seo, Q. Xu, H. W. Zandbergen, A. Yazdani, N. P. Ong, and R. J. Cava, *Superconductivity in CuxBi2Se3 and Its Implications for Pairing in the Undoped Topological Insulator*, Phys Rev Lett **104**, 057001 (2010).

[8] S. Ran et al., *Nearly Ferromagnetic Spin-Triplet Superconductivity*, Science (1979) **365**, 684 (2019).

[9] K. E. Avers et al., *Broken Time-Reversal Symmetry in the Topological Superconductor UPt3*, Nat Phys **16**, 531 (2020).

[10] Y. Ando and L. Fu, *Topological Crystalline Insulators and Topological Superconductors: From Concepts to Materials*, Annu. Rev. Condens. Matter Phys **6**, 361 (2015).

[11] S. Sasaki, Z. Ren, A. A. Taskin, K. Segawa, L. Fu, and Y. Ando, *Odd-Parity Pairing and Topological Superconductivity in a Strongly Spin-Orbit Coupled Semiconductor*, Phys Rev Lett **109**, 217004 (2012).

[12] L. Fu, *Topological Crystalline Insulators*, Phys Rev Lett **106**, 106802 (2011).

[13] T. H. Hsieh, H. Lin, J. Liu, W. Duan, A. Bansil, and L. Fu, *Topological Crystalline Insulators in the SnTe Material Class*, Nat Commun **3**, 982 (2012).

[14] M. Novak, S. Sasaki, M. Kriener, K. Segawa, and Y. Ando, *Unusual Nature of Fully Gapped Superconductivity in In-Doped SnTe*, RAPID COMMUNICATIONS PHYSICAL REVIEW B **88**, 140502 (2013).

[15] P. M. R. Brydon, S. Das Sarma, H.-Y. Hui, and J. D. Sau, *Odd-Parity Superconductivity from Phonon-Mediated Pairing: Application to CuxBi2Se3*, Phys Rev B **90**, 184512 (2014).

[16] C. M. Polley et al., *Observation of Surface States on Heavily Indium-Doped SnTe(111), a Superconducting Topological Crystalline Insulator*, Phys Rev B **93**, 075132 (2016).

[17] G. Balakrishnan, L. Bawden, S. Cavendish, and M. R. Lees, *Superconducting Properties of the In-Substituted Topological Crystalline Insulator SnTe*, Phys Rev B **87**, 140507 (2013).

[18] M. P. Smylie et al., *Superconductivity, Pairing Symmetry, and Disorder in the Doped Topological Insulator Sn1-XInxTe for X>0.1*, Phys Rev B **97**, 024511 (2018).

[19] T. Nomoto, M. Kawamura, T. Koretsune, R. Arita, T. Machida, T. Hanaguri, M. Kriener, Y. Taguchi, and Y. Tokura, *Microscopic Characterization of the Superconducting Gap Function in Sn1-x InxTe*, Phys Rev B **101**, 014505 (2020).

[20] C. Zhang, X.-G. He, H. Chi, R. Zhong, W. Ku, G. Gu, J. M. Tranquada, and Q. Li, *Electron and Hole Contributions to Normal-State Transport in the Superconducting System ${\rm Sn}_{1-X}{\rm In}_x{\rm Te}$*, Phys Rev B **98**, 054503 (2018).

[21] N. Haldolaarachchige, Q. Gibson, W. Xie, M. B. Nielsen, S. Kushwaha, and R. J. Cava, *Anomalous Composition Dependence of the Superconductivity in In-Doped SnTe*, Phys Rev B **93**, 024520 (2016).

[22] A. S. Erickson, J.-H. Chu, M. F. Toney, T. H. Geballe, and I. R. Fisher, *Enhanced Superconducting Pairing Interaction in Indium-Doped Tin Telluride*, Phys Rev B **79**, 024520 (2009).

[23] J. Wang et al., *Observation of Coexisting Weak Localization and Superconducting Fluctuations in Strained Sn 1– x In x Te Thin Films*, Nano Lett **22**, 792 (2022).

[24] W. Si, C. Zhang, L. Wu, T. Ozaki, G. Gu, and Q. Li, *Superconducting Thin Films of (100) and (111) Oriented Indium Doped Topological Crystalline Insulator SnTe*, Appl Phys Lett **107**, 092601 (2015).

[25] M. Masuko, R. Yoshimi, A. Tsukazaki, M. Kawamura, K. S. Takahashi, M. Kawasaki, and Y. Tokura, *Molecular Beam Epitaxy of Superconducting Sn1-XInxTe Thin Films*, Phys Rev Mater **4**, 091202 (2020).

[26] N. Haldolaarachchige, Q. Gibson, W. Xie, M. B. Nielsen, S. Kushwaha, and R. J. Cava, *Anomalous Composition Dependence of the Superconductivity in In-Doped SnTe*, Phys Rev B **93**, 024520 (2016).

[27] P. Makuła, M. Pacia, and W. Macyk, *How To Correctly Determine the Band Gap Energy of Modified Semiconductor Photocatalysts Based on UV–Vis Spectra*, J Phys Chem Lett **9**, 6814 (2018).

[28] H. Preier, *Recent Advances in Lead-Chalcogenide Diode Lasers*, Applied Physics **20**, 189 (1979).

[29] S. Rabii, *Energy-Band Structure and Electronic Properties of SnTe*, Physical Review **182**, 821 (1969).

[30] T. C. Chasapis, Y. Lee, E. Hatzikraniotis, K. M. Paraskevopoulos, H. Chi, C. Uher, and M. G. Kanatzidis, *Understanding the Role and Interplay of Heavy-Hole and Light-Hole Valence Bands in the Thermoelectric Properties of PbSe*, Phys Rev B **91**, 085207 (2015).

[31] J. He et al., *Valence Band Engineering and Thermoelectric Performance Optimization in SnTe by Mn-Alloying via a Zone-Melting Method*, J Mater Chem A Mater **3**, 19974 (2015).

[32] I. I. Zasavitskii, E. A. de Andrada e Silva, E. Abramof, and P. J. McCann, *Optical Deformation Potentials for PbSe and PbTe*, Phys Rev B **70**, 115302 (2004).

[33] X. Zhang and S. Ptasinska, *Dissociative Adsorption of Water on an H2O/GaAs(100) Interface: In Situ Near-Ambient Pressure XPS Studies*, The Journal of Physical Chemistry C **118**, 4259 (2014).

[34] V. S. Neudachina, T. B. Shatalova, V. I. Shtanov, L. V. Yashina, T. S. Zyubina, M. E. Tamm, and S. P. Kobeleva, *XPS Study of SnTe(100) Oxidation by Molecular Oxygen*, Surf Sci **584**, 77 (2005).

[35] C. M. Varma, *Missing Valence States, Diamagnetic Insulators, and Superconductors*, Phys Rev Lett **61**, 2713 (1988).

[36] P. Walmsley, C. Liu, A. D. Palczewski, P. Giraldo-Gallo, C. G. Olson, I. R. Fisher, and A. Kaminski, *Direct Spectroscopic Evidence for Mixed-Valence Tl in the Low Carrier-Density Superconductor Pb1-XTlxTe*, Phys Rev B **98**, 184506 (2018).

[37] M. Matusiak, E. M. Tunnicliffe, J. R. Cooper, Y. Matsushita, and I. R. Fisher, *Evidence for a Charge Kondo Effect in Pb1-XTlxTe from Measurements of Thermoelectric Power*, Phys Rev B **80**, 220403 (2009).

[38] G. E. McGuire, G. K. Schweitzer, and T. A. Carlson, *Core Electron Binding Energies in Some Group IIIA, VB, and VIB Compounds*, Inorg Chem **12**, 2450 (1973).

[39] R. D. Zhong, J. A. Schneeloch, X. Y. Shi, Z. J. Xu, C. Zhang, J. M. Tranquada, Q. Li, and G. D. Gu, *Optimizing the Superconducting Transition Temperature and Upper Critical Field of Sn1-XInxTe*, Phys Rev B **88**, 020505 (2013).

[40]  S. Misra, B. Wiendlocha, J. Tobola, P. Levinský, J. Hejtmánek, S. Migot, J. Ghanbaja, A. Dauscher, B. Lenoir, and C. Candolfi, *Influence of In-Induced Resonant Level on the Normal-State and Superconducting Properties of Sn1.03Te*, Phys Rev B **106**, 075205 (2022).

[41]  B. H. Freeland, J. J. Habeeb, and D. G. Tuck, *Coordination Compounds of Indium. Part XXXIII. X-Ray Photoelectron Spectroscopy of Neutral and Anionic Indium Halide Species*, Can J Chem **55**, 1527 (1977).